\documentclass[aps,prl,twocolumn,nofootinbib,
   showpacs,
   superscriptaddress]{revtex4}
\usepackage{graphicx}
\usepackage{amsmath}
\usepackage{amssymb}

\begin{document}

\title{\boldmath Possible Solution of the  $J/\psi$ Production Puzzle}

\author{H. Haberzettl}
\affiliation{\mbox{Center for Nuclear Studies,
 Department of Physics, The George
Washington University, Washington, DC 20052, USA}}

\author{J.\,P. Lansberg}
\affiliation{Institut f\"ur Theoretische Physik,
 Universit\"at Heidelberg, 69120 Heidelberg, Germany}
\affiliation{Centre de Physique Th\'eorique,
 \'Ecole Polytechnique, CNRS, 91128 Palaiseau, France}

\begin{abstract}
We argue that the $s$-channel cut contribution to $J/\psi$ hadroproduction can
be significantly larger than the usual cut contribution of the color-singlet
mechanism (CSM), which is known to underestimate the experimental measurements.
A scenario accounting for intermediate $c\bar{c}$ interactions is proposed that
reproduces the data at low- and mid-range transverse momenta $P_T$ from the
Fermilab Tevatron and BNL Relativistic Heavy Ion Collider. The $J/\psi$
produced in this manner are polarized predominantly longitudinally.
\end{abstract}

\pacs{13.60.Le, 11.40.-q, 13.85.Ni,  14.40.Gx
     \hfill PRL \textbf{100}, 032006 (2008)---arXiv:0709.3471v2
   }
 \maketitle

Although heavy quarkonia  are among the most analyzed bound quark systems, ever
since the first measurements by the CDF Collaboration of the {\it direct}
production of $J/\psi$ and $\psi'$ at $\sqrt{s}=1.8$
TeV~\cite{Abe:1997jz,Abe:1997yz} we are facing persistent disagreements between
theoretical predictions from various models and experimental studies of the
cross section and the polarization (for reviews see~\cite{review}).

The recent confirmation by CDF~\cite{Abulencia:2007us} of their previous
polarization measurement~\cite{Affolder:2000nn} showing an unpolarized or
slightly longitudinally polarized $J/\psi$ yield  has reinforced doubts about
the applicability of the quark-velocity expansion ($v$) of nonrelativistic QCD
(NRQCD)~\cite{Bodwin:1994jh} for the  rather ``light'' $c\bar c$ system. On the
theory side, many new results completed --- but also questioned
--- our knowledge of charmonium production: the long-awaited next-to-leading
order (NLO) QCD corrections to the color-singlet
contributions~\cite{Campbell:2007ws} are now available and show significant
enhancement of the cross section; an up-to-date proof~\cite{Nayak1} of NRQCD
factorization holding true at any order in $v$ in the gluon-fragmentation
channels was provided; the universality of the nonperturbative input of NRQCD
was challenged by fixed-target measurements~\cite{Maltoni:2006yp}, similar to
what had been found previously for the HERA data~\cite{review}; NRQCD
factorization was shown to require modification in fragmentation regions where
three heavy quarks have similar momenta~\cite{Nayak:2007mb}; finally, the
$c$-quark fragmentation approximation was shown~\cite{Artoisenet:2007xi} to be
only valid at much higher $P_T$ than expected from the pioneering works of
Ref.~\cite{frag_CSM}.

\begin{figure}[b!]\centering
  \includegraphics[width=.85\columnwidth,clip=]{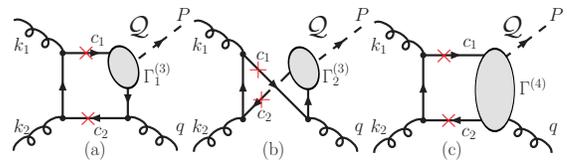}
  \caption{\label{fig:Jpsibox}%
  (Color online).
  (a,b) Leading-order (LO) $s$-channel cut diagrams contributing to $gg\to
\mathcal{Q}g$ with direct and crossed box diagrams employing  the
$c\bar{c}\,J/\psi$ vertex.  The crosses indicate that the quarks are on-shell.
(c) Box diagram with $c\bar{c}\mathcal{Q}g$ contact term mandated by gauge
invariance. }
\end{figure}

In view of the difficulty of most theoretical approaches in reproducing the
experimental data, the authors of Ref.~\cite{Lansberg:2005pc} undertook a
systematic study of the cut contributions due to off-shell and nonstatic
quarks. In particular, they questioned the assumption of the CSM that takes the
heavy quarks forming the quarkonium ($\mathcal{Q}$) as being
on-shell~\cite{CSM_hadron}. If they are not, the usual $s$-channel cut
contributes to the imaginary part of the amplitude and need to be considered on
the same footing as the CSM cut.  These  $s$-channel cut contributions, which
were the specific focus in Ref.~\cite{Lansberg:2005pc}, are illustrated in
Fig.~\ref{fig:Jpsibox}.

\begin{figure}[t!]\centering
\includegraphics[width=.6\columnwidth,clip=]{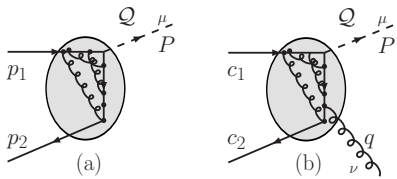}
 \caption{\label{fig:illus_GI_break}%
Example of the mechanisms (a) contributing to the dressing of the 3-point
function $\Gamma^{(3)}$ and (b) responsible for the 4-point function
$\Gamma^{(4)}$: the external gluon is attached here within gluon loops of the
dressed vertex, thus producing a 
diagram without poles
and with a kinematic behavior genuinely different from the initial 3-point
vertex.}
\end{figure}

In order to provide a conserved current for such off-shell configurations, one
must introduce an additional four-point function, or contact
current~\cite{Drell:1971vx}. Dynamically such a current accounts for the
interactions between the $c\bar{c}$ pair emitting the external gluon. In fact,
this mechanism arises because of the possibility that the outgoing gluon stems
from \emph{within} the dressed $c\bar{c}\,J/\psi$ vertex, as depicted in
Fig.~\ref{fig:illus_GI_break}. The quark pair $(c_1,c_2)$ that makes the final
$J/\psi$-gluon state is now in a color-octet state which thus recovers the
necessity for such configurations as a natural consequence of restoring gauge
invariance.

The construction of this 4-point function based on gauge invariance alone is
not unique since additional transverse contributions have no bearing on current
conservation. In principle,  the details of this 4-point function and of the
necessary transverse contributions would follow from a full dynamical treatment
that consistently accounts for all such interactions. At present at least, this
is beyond the scope of what is theoretically possible.%

Yet, it is possible to construct a 4-point function that satisfies certain
minimal requirements~\cite{Kazes:1959,Drell:1971vx}. The 4-point function
proposed in~\cite{Lansberg:2005pc} provided a conserved current but was not
entirely satisfactory since it contained  poles (similar to the basic direct
and crossed contributions), and such poles for the contact current are
unphysical and therefore should be avoided~\cite{Drell:1971vx}.

To this end, drawing on the experience gained in pion photoproduction
processes~\cite{Haberzettl:1997jg,Haberzettl:1998eq,Davidson:2001rk,Haberzettl:2006fsi},
we propose here a contact current for $J/\psi$ production which satisfies the
requirements of full gauge invariance, going beyond mere current conservation,
in terms of the generalized Ward--Takahashi identity
\cite{Kazes:1959,Haberzettl:1997jg}. The basic mechanism how to do this was
proposed a long time ago by Drell and Lee~\cite{Drell:1971vx} for the
electromagnetic case, where it basically amounts to the minimal substitution
prescription $\partial^\mu \to \partial^\mu +i Q A^\mu$  ($Q$: charge, $A^\mu$:
vector potential) in an effective Lagrangian corresponding to the dressed
hadronic vertex. As noted by Drell and Lee, this particular approach is
deficient in that it violates the high-energy scaling behavior, because in
avoiding poles for the 4-point function, it partially replaces the true
momentum dependence of the vertices by constants. We avoid this shortcoming by
building a phenomenological vertex incorporating this low-energy behavior and
the expected momentum dependence of the vertices at high energies by
interpolating between these two regimes. As we show below, the resulting
enhanced $s$-channel cut contributions reproduce the experimental data from
both the CDF and PHENIX collaborations up to transverse momenta $P_T\simeq 10$
GeV.

The present formalism builds on the approach developed
in~\cite{Lansberg:2005pc}, where the transition $q\bar q\rightarrow {\mathcal
Q}$ is described by the 3-point function
\begin{equation}\label{vf}
\Gamma^{(3)}_{\mu}(p,P) = \Gamma(p,P) \gamma_\mu~,
\end{equation}
where  $P\equiv p_{1}-p_{2}$  and $p\equiv(p_{1}+p_{2})/2$ are the total and
relative momenta, respectively, of the two quarks bound as a quarkonium state,
with $p_1$ and $p_2$ being their individual four-momenta. Ansatz (\ref{vf})
amounts to representing the vector meson as a massive photon with a nonlocal
coupling. In the present work, we describe the relative-momentum distribution
$\Gamma(p,P)$ of the quarks phenomenologically as a Gaussian, as in
Ref.~\cite{Lansberg:2005pc}, where full details can be found.

The generic picture of the physical origin of the dressed vertex function
$\Gamma(p,P)$ is given in Fig.~\ref{fig:illus_GI_break}(a), and the ensuing
contact current when coupling the gauge boson \emph{into} this vertex is
illustrated in Fig.~\ref{fig:illus_GI_break}(b).

The requirement of gauge invariance is usually written in terms of the
generalized Ward--Takahashi relations~\cite{Kazes:1959} for the complete
current. For the purpose of restoring gauge invariance, it is most convenient
to rewrite this into an equivalent condition for the contact current (see
\cite{Haberzettl:1997jg} for the analogous procedure in pion photoproduction).
To this end, let us write the 4-point function depicted in
Fig.~\ref{fig:illus_GI_break}(b) as
\begin{equation}
\Gamma^{(4)} =-ig_s T^{a}_{ik} M_c^\nu \gamma^\mu~,
\label{eq:4pointfct}
\end{equation}
 where $g_s$ is the strong coupling constant, $T^{a}_{ik}$ the
color matrix, and $\mu$ and $\nu$ are the Lorentz indices of the outgoing
$J/\psi$ and gluon, respectively. For simplicity, we have suppressed all
indices on the left-hand side. The $c\bar{c}\,J/\psi$ vertex function
$\Gamma^{(3)}$ with the kinematics of the direct graph is denoted here by
$\Gamma_1$ and for the crossed graph by $\Gamma_2$, {\it i.e.}, $\Gamma_1 =
\Gamma\left(c_1-\frac{P}{2},P\right)$ and $\Gamma_2 =
\Gamma\left(c_2+\frac{P}{2},P\right)$, as shown in Figs.~\ref{fig:Jpsibox}(a)
and \ref{fig:Jpsibox}(b). The gauge-invariance condition for the contact
current $M_c^\nu$ for the outgoing gluon with momentum $q$ reads now
\begin{equation}
q_\nu M_c^{\nu} = \Gamma_1-\Gamma_2
 \label{gipcond}
\end{equation}
since this is precisely the four-divergence contribution needed to cancel the
corresponding terms arising from the four-divergences of
Figs.~\ref{fig:Jpsibox}(a) and \ref{fig:Jpsibox}(b).

We emphasize here that, within the present semi-phenomenological approach, the
procedure to preserve QCD gauge invariance for the gluon coupling follows
exactly along the lines of QED for a photon since the four-point
function~(\ref{eq:4pointfct}) factorizes in the color matrix and the gluon
coupling. This finding is directly related to the fact that the two vertex
functions on the right-hand side of the four-divergence contribution
(\ref{gipcond}) arise from the gluon coupling to the two intermediate quark
lines in Figs.~\ref{fig:Jpsibox}(a) and \ref{fig:Jpsibox}(b) which, apart from
color factors, is exactly like a photon coupling to spin-1/2 particles.

We now employ the usual construction
\cite{Haberzettl:1997jg,Haberzettl:1998eq,Davidson:2001rk,Haberzettl:2006fsi}
for the contact current in terms of an auxiliary function $F=F(c_1,c_2,q)$ and
put
\begin{equation}
M_c^\nu = \frac{(2c_2+q)^\nu\left(\Gamma_1-F\right)}{(c_2+q)^2-m^2}
 +\frac{(2c_1-q)^\nu\left(\Gamma_2-F\right)}{(c_1-q)^2-m^2}~,
 \label{Mc}
\end{equation}
where we take  $c_1^2=c_2^2=m^2$ and $P^2=M^2$ from the beginning, with $m$ and
$M$ being the masses of the quark and the  $J/\psi$, respectively. One easily
verifies that this additional contact current satisfies the gauge-invariance
condition (\ref{gipcond}). It is found, in particular, that $F$ cancels out in
the four-divergence.

The function $F(c_1,c_2,q)$ must be chosen so that the current (\ref{Mc})
satisfies crossing symmetry (\textit{i.e.}, symmetry under the exchange
$c_1\leftrightarrow -c_2$) and is free of singularities. The latter constraint
implies $F=\Gamma_0$ at either pole position, {\it i.e.}, when $(c_2+q)^2=m^2$
or $(c_1-q)^2=m^2$, where the constant $\Gamma_0$ is the (unphysical) value of
the momentum distribution $\Gamma(p,P)$ when all three legs of the vertex are
on their respective mass shells. In principle, employing gauge invariance as
the only constraint, we may take $F=\Gamma_0$ everywhere. This corresponds to
the minimal substitution discussed by Drell and Lee~\cite{Drell:1971vx} (for a
complete derivation see~\cite{Ohta:1989ji}) who pointed out, however, that this
does not provide the correct scaling properties at large energies, which means
within the present context that $F=\Gamma_0$ would not lead to the expected
$P_T$ scaling of the amplitude. Numerically, this choice overshoots the
experimental data by more than one order of magnitude at $P_T=20$ GeV. In fact,
in the large relative-momentum region, we expect the contact term and therefore
the function $F(c_1,c_2,q)$ to exhibit a fall-off similar to the vertex
functions themselves, contrary to the minimal substitution procedure. The
simplest crossing-symmetric choice with this behavior is $F= \Gamma_1+\Gamma_2
-{\Gamma_1\Gamma_2}/{\Gamma_0}$ ~\cite{Davidson:2001rk}.
The solution we propose here is to build $F(c_1,c_2,q)$ from these two limiting
cases. To this end, it is natural to choose the following simple ansatz
\begin{equation}
F(c_1,c_2,q)= \Gamma_0-h(c_1
c_2)\frac{\left(\Gamma_0-\Gamma_1\right)\left(\Gamma_0-\Gamma_2\right)}{\Gamma_0}~,
\end{equation}
where the (crossing-symmetric) function $h(c_1 c_2)$  rises to become unity for
large relative momentum. Note that it is not necessary that $h(c_1 c_2)$
actually vanish at either pole since the factor on its right vanishes for
$(c_2+q)^2=m^2$ or $(c_1-q)^2=m^2$ and we recover the choice of $F=\Gamma_0$ at
either pole. The phenomenological choice for the inter\-polating function
$h(c_1 c_2)$ used in our calculations is
\begin{equation}
h(c_1 c_2)= 1- a\frac{\kappa^2}{\kappa^2-(c_1 c_2+m^2)}~,
 \label{eq:interpolate}
\end{equation}
with two parameters, $a$ and $\kappa$. We would like to emphasize at this point
that this choice is in no way unique. In a manner of speaking, this choice is
simply a way of parameterizing our ignorance by employing minimal properties of
$\Gamma^{(4)}$. We shall not, however, discuss other choices here since our
main motivation is to show that $s$-channel cut contributions can be large and
can indeed reproduce the data --- the physical picture that emerges could then
be tested in other production regimes.

In the kinematical region accessed at the Tevatron, the direct $J/\psi$ are
produced by gluon fusion and a final-state gluon emission is required to
conserve $C$-parity and provide the $J/\psi$ with its $P_T$. The relevant
diagrams for the  LO gluon fusion process can be found
in~\cite{Lansberg:2005pc}, the only difference in the present treatment being
the new choice of $\Gamma^{(4)}$. Also, we use the same normalization of
$\Gamma^{(3)}$ as in~\cite{Lansberg:2005pc}.


\begin{figure}[b!]\centering
\includegraphics[width=.73\columnwidth,clip=]{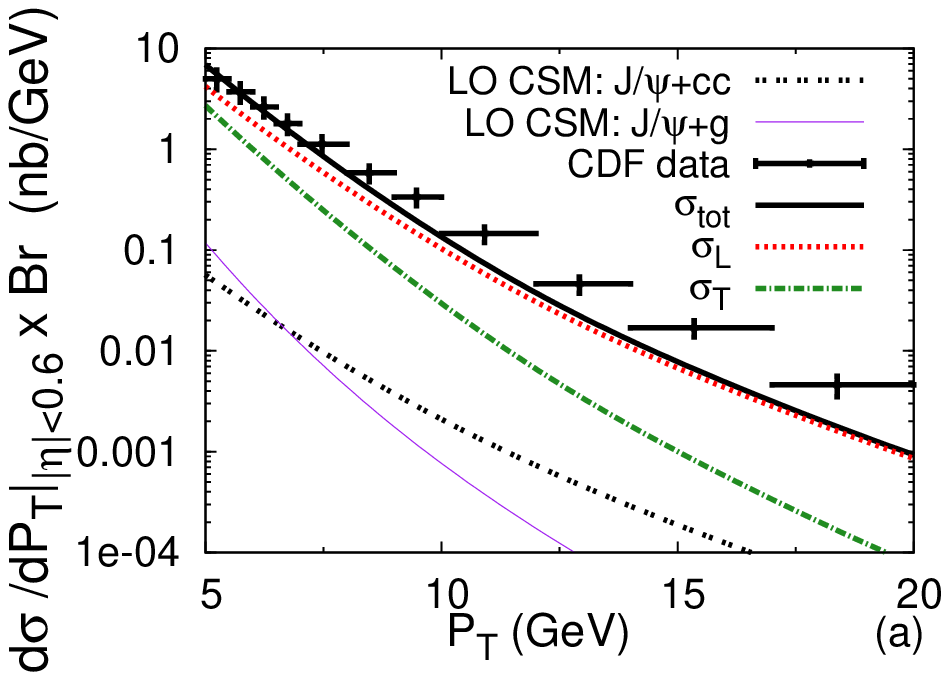}\\[2ex]
\includegraphics[width=.73\columnwidth,clip=]{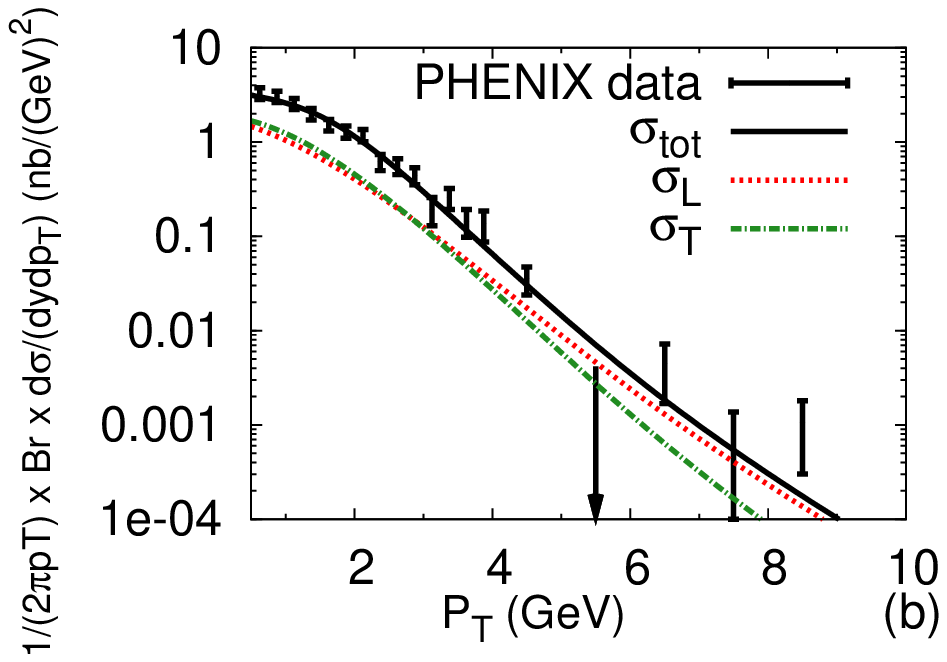}
\caption{(Color online).
(a) Comparison between polarized ($\sigma_T$ and $\sigma_L$) and
unpolarized ($\sigma_{\text{tot}}$) cross sections [with parameters $a=4$,
$\kappa=4.5$ GeV in Eq.~(\ref{eq:interpolate})], LO CSM contributions, and CDF
experimental data~\cite{Abe:1997yz} at the Tevatron ($\sqrt{s}=1.8$ TeV,
pseudorapidity $|\eta|<0.6$). (b) Comparison between $\sigma_T$, $\sigma_L$,
$\sigma_{\text{tot}}$ and PHENIX data~\cite{Adare:2006kf} at RHIC
($\sqrt{s}=200$ GeV, rapidity $|y|<0.35$).} \label{fig:cross-sections}
\end{figure}

The double-differential polarized cross section in transverse momentum $P_T$
and rapidity $y$ is given by~\cite{Lansberg:2005pc}
\begin{equation}
\frac{d\sigma_r}{dy\,dP_T}=\int_{x_1^{\text{min}}}^1 dx_1 \frac{2 \hat s P_T
g(x_1) g\left(x_2(x_1)\right)} {\sqrt{s}(\sqrt{s}x_1-E_T
e^{y})}\frac{d\sigma_r}{d\hat t}~,
\end{equation}
where ${d\sigma_r}/{d\hat{t}}$ is the  partonic differential cross section,
with $r=L,T_1,T_2$ being the quarkonium helicity, and $\hat s=(k_1+k_2)^2$,
$\hat t=(k_2-q)^2$ and $\hat u=(k_1-q)^2$ are the Mandelstam variables for the
partonic process.  In the present calculations, we use the LO gluon
distribution $g(x)$ of~\cite{Martin:2002dr}, and the same mass and size
parameter $\Lambda$ as given in Ref.~\cite{Lansberg:2005pc}. In any case, our
conclusion that the $s$-channel cut contribution can reproduce the experimental
data with adjustments of the values for $a$ and $\kappa$ would not change at
all over a wide range for these parameters.

Figure~\ref{fig:cross-sections}(a) shows our results with parameter values
$a=4$ and $\kappa=4.5$\,GeV for $\sqrt{s}=1.8$ TeV in the pseudorapidity range
$|\eta|<0.6$ compared with the cross-section measurement of direct $J/\psi$ by
CDF~\cite{Abe:1997yz}, the usual LO CSM from $gg\to ~J/\psi\,
g$~\cite{CSM_hadron} and LO CSM from $gg\to ~J/\psi\, c
\bar{c}$~\cite{Artoisenet:2007xi}. Our results agree very well with the CDF
data up to about $P_T=10$ GeV. At higher $P_T$, our curve falls below the data
as expected from the genuine ${1}/{P^8_T}$ scaling of a LO box diagram.
Inclusion of higher-order corrections incorporating fragmentation-type
topologies ($\sim{1}/{P^4_T}$) and associated-production channels are expected
to fill the gap between data and theory at high $P_T$.
It is interesting to note the different $P_T$ behaviors of $\sigma_T$ and
$\sigma_L$ leading to a dominance of the latter at large $P_T$ and a negative
value for the polarization $\alpha$. Figure~\ref{fig:cross-sections}(b) shows
our results at $\sqrt{s}=200$\,GeV, still with  $a=4$ and $\kappa=4.5$\,GeV,
compared with the PHENIX data~\cite{Adare:2006kf}.

\begin{figure}[t!]\centering
\includegraphics[width=.85\columnwidth,clip=true]{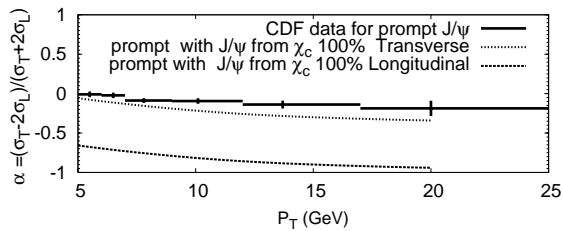}
\caption{Prompt $J/\psi$ polarization: theory \textit{vs.} CDF
data~\cite{Abulencia:2007us}.} \label{fig:polarisation}
\end{figure}

Since $J/\psi$ polarization measurements exist only for the prompt yield, we
have computed $\alpha$ from our direct-$J/\psi$ cross sections in two extreme
cases, one where the $J/\psi$'s from $\chi_c$ are 100\% transverse and another
where they are 100\% longitudinal, the first scenario being the more likely
one. Figure~\ref{fig:polarisation} shows the comparison between this
computation and the recent results by CDF at $\sqrt{s}=$ 1.96
TeV~\cite{Abulencia:2007us}.

In conclusion, we have shown here that among the two singularities in the box
diagram contribution to quarkonium production at LO, the one from the
$s$-channel cut is significantly larger than expected when one includes
interactions between the quark pair which binds into the quarkonium and emits
the final-state gluon. In the present work, we have accounted for the effect of
those interactions by a phenomenological contact current, or 4-point function,
$\Gamma^{(4)}$ which follows directly from the implementation of the full
gauge-invariance requirements appropriate for a dressed vertex $\Gamma^{(3)}$.
This 4-point function, however, is not fully constrained by gauge invariance
which permitted us to interpolate between the minimal substitution discussed
earlier by Drell and Lee and the expected large relative-momentum behavior.

As we showed, this 4-point function provides a much larger contribution than
the direct and crossed ones containing 3-point functions (cf.\
Fig.~\ref{fig:Jpsibox}), and this can easily bring about agreement with the
experimental data. In NRQCD, color-octet matrix elements account for
transitions between a colored heavy-quark pair into a quarkonium by soft unseen
gluon emission in the final state. In the present approach, the 4-point
function accounts for gluon exchanges between the heavy quarks which emit the
final-state gluon. As for the matrix elements of NRQCD, which are unknown and
then fit, we fixed the unconstrained parameter of this function in order to
reproduce the experimental data at $\sqrt{s}=1.8$ TeV from the CDF
collaboration at the Tevatron for $P_T\lesssim 10$GeV.

With the same parameters, we also obtain a very good description of the
experimental measurements from PHENIX at RHIC at $\sqrt{s}=200$ GeV. Moreover,
our prediction for the polarization for the prompt  $J/\psi$  yield is mostly
longitudinal. This looks promising since one expects contributions from the CSM
cuts --- which are known to be enhanced by the NLO
corrections~\cite{Campbell:2007ws} --- and from the real part which is not
evaluated at present time.

A similar enhancement by inclusion of the $s$-channel cut is expected in all
production processes where the $J/\psi$ is associated with a gluon,
\textit{e.g.},~photon-photon collision at LEP as well as in photo- and
lepto-production at HERA.

We thank P. Artoisenet, G. Bodwin, E. Braaten, S.\,J. Brodsky, J. Collins,
J.\,R. Cudell, Yu.\,L. Kalinovsky, M.\,J. Kim, J.\,W. Qiu, F. Maltoni, V.
Papadimitriou, T.\,N. Pham, and B. Pire for useful discussions. The work of
J.\,P.\,L. is supported by the European contract RII3-CT-2004-506078.

\end{document}